\documentclass[aps, reprint,prl,superscriptaddress,nofootinbib]{revtex4-2}
\usepackage{dcolumn}
\usepackage[utf8]{inputenc}
\usepackage{amsmath}
\usepackage{amsfonts}
\usepackage{amssymb}
\usepackage{mathtools}
\usepackage{graphicx}
\usepackage{hyperref}
\usepackage{xcolor}
\usepackage{comment}
\usepackage[toc,page]{appendix}
\usepackage[english]{babel}
\newtheorem{teorema}{Principle}

\begin{document}

\title{On the Role of Locality in the Bose-Marletto-Vedral Effect}

\author{Giuseppe Di Pietra}
\affiliation{Clarendon Laboratory, University of Oxford, Parks Road, Oxford OX1 3PU, United Kingdom}

\author{Vlatko Vedral}
\affiliation{Clarendon Laboratory, University of Oxford, Parks Road, Oxford OX1 3PU, United Kingdom}

\author{Chiara Marletto}
\email{chiara.marletto@gmail.com}
\affiliation{Clarendon Laboratory, University of Oxford, Parks Road, Oxford OX1 3PU, United Kingdom}

\date{\today}%

\begin{abstract}
Two of us recently proposed an entanglement-based witness of non-classicality, which can be applied to testing quantum effects in gravity in what is known as the Bose-Marletto-Vedral (BMV) effect. The witness is based on this idea: if a system can create entanglement between two quantum probes by local means only, then it must be non-classical. In this note we discuss the role of locality as an assumption for the theorem supporting the witness; we also discuss other related notions of locality and comment on their mutual relations. 
\end{abstract}

\maketitle


\section{Introduction}
\begin{figure}
    \centering
    \includegraphics[width=\columnwidth]{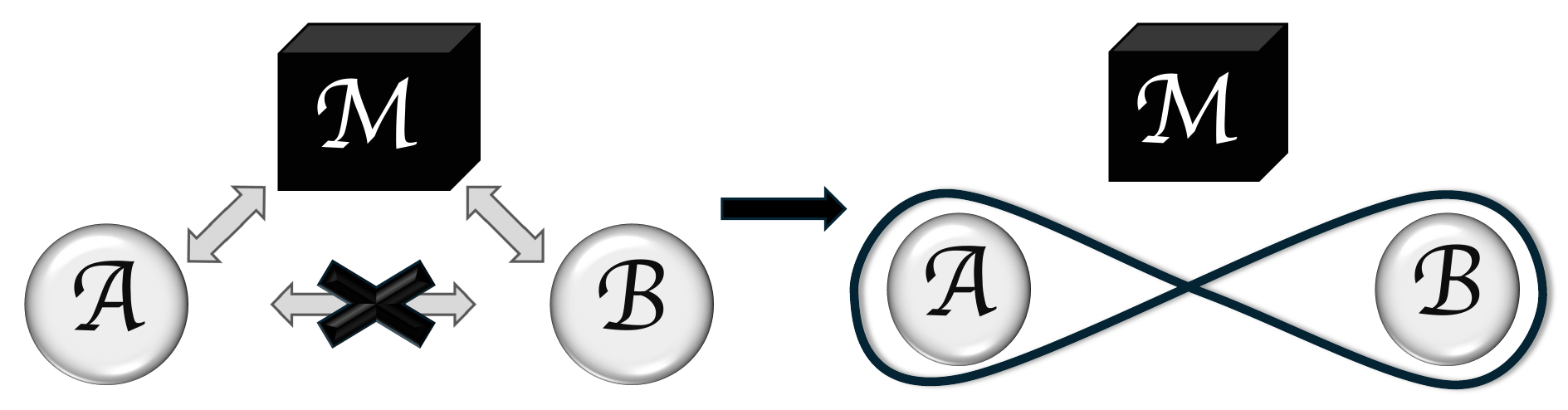}
    \caption{Schematic representation of the setup for the General Witness Theorem. The two space-like separated quantum probes $A$ and $B$ are coupled only via the unknown system $M$, by means of local interactions. Its capability of inducing entanglement between $A$ and $B$ would then be a witness of its non-classicality.}
    \label{fig:BMV}
\end{figure}

A particularly promising approach to testing quantum gravity has recently been proposed, based on a novel ``witness of non-classicality''. This witness relies on the entangling power of a given system to conclude that the system has non-classical features. In particular, these tests are based on the so-called General Witness Theorem (GWT), \cite{marletto_quantum-information_2024,marletto_witnessing_2020}, stating that if a system $M$ (such as gravity) can mediate (by local means) entanglement between two quantum systems, $A$ and $B$, (e.g. two masses) then it must be non-classical, \cite{marletto_witnessing_2020}. By ``local means" here we mean a specific protocol, detailed in \cite{marletto_witnessing_2020, bose_spin_2017, marletto_gravitationally_2017}, where $A$ and $B$ must not interact directly with each other, but only via the mediator $M$, as schematically represented in Fig.\ref{fig:BMV}. Interestingly, ``non-classicality'' is a theory-independent generalisation of what in quantum theory is expressed as ``having at least two distinct physical variables that do not commute", which can be expressed within a general information-theoretic framework, the constructor theory of information, \cite{deutsch_constructor_2015}. Informally, being non-classical means having two or more distinct physical variables that cannot simultaneously be measured to an arbitrarily high degree of accuracy, \cite{marletto_witnessing_2020}. Due to its generality, the GWT offers a broad theoretical basis for recently proposed experiments that can test quantum effects in gravity at the laboratory scale, based on the generation of gravitational entanglement between two massive probes - the so-called Bose-Marletto-Vedral effect, \cite{bose_spin_2017,marletto_gravitationally_2017}. It also provides a basis for any other experiment that (beyond the case of gravity) intends to show that some system $M$ is non-classical \cite{raia_role_2024}, using the effect of its entangling power. 

A particularly appealing feature of the GWT is that, by using the constructor theory of information, it avoids assuming the usual machinery of quantum information theory, thus extending beyond quantum theory existing results such as the theorems that forbid the creation of entanglement via local operations and classical communication. Moreover, the GWT is proven without assuming the existence of a probability space, in contrast to existing approaches such as Generalised Probabilistic Theories \cite{plavala_general_2023}. This generality is particularly important as one wants to use it in a context where the system $M$ may or may not obey quantum theory itself. 

This witness relies on the capacity of a system $M$ to generate entanglement between two independent subsystems, initially unentangled with each other. For the witness to be applicable, it is key that the systems $A$ and $B$ are independent -- hence it is essential to assume the \textit{principle of locality}. We shall discuss here 1) what the minimal notion of locality on which the witness relies is; 2) how the principle of locality is different and more general than other notions of locality in physics; and 3) what known theories satisfy this principle, and hence what the implications of the witness are for those theories. We shall also briefly highlight how the principle of locality is used in the proof of the GWT.

\section{Summary of the witness of non-classicality}

In this Section, we recall the key idea of the witness and explain, using an example from quantum theory, how the witness works to conclude that a system is non-classical. 
A witness of non-classicality is a protocol to probe a system $M$, whose dynamics is partly unknown, with one or more fully quantum probes $Q$, to the end of establishing whether $M$ has some quantum features by measuring only the quantum probes. 

The scenario we have in mind is one in which $M$ has dynamics that could be fully classical (as in the case of gravity) or affected by dynamical collapse (as in the case of a macroscopic object, e.g., a complex biomolecule). 

The witness we shall focus on is the one where there are two quantum probes, $A$ and $B$, which interact via $M$ -- which is a mediator. If one can set up an experiment where the probes interact via $M$ only, and manage to get entangled via $M$, then the GWT allows one to conclude that $M$ is non-classical. 

As we mentioned, the GWT can be proven in a very general framework, without assuming quantum theory's formalism. In this Section, however, we shall give a specific example of how the theorem works, using quantum theory. In this example, $M$ being non-classical means that it has at least two variables that do not commute.

{Let us assume, for simplicity, that $A$ and $B$ are two qubits. Since our goal is to focus on the role of the principle of locality in this witness, we shall describe the qubits using the descriptors formalism, which stems from the Heisenberg picture of quantum theory \cite{deutsch_information_2000,bedard_abc_2021}. In this formalism, one describes the qubit $A$ with a vector of \textit{descriptors}: 
\begin{align}
    \hat{q}^{(A)}(t_0)&\coloneqq \left(q_x^{(A)}(t_0),q_y^{(A)}(t_0),q_z^{(A)}(t_0)\right) \nonumber & \\ & =\left(\sigma_x \otimes \mathrm{I}_M \otimes \mathrm{I}_B,\sigma_y\otimes \mathrm{I}_M \otimes \mathrm{I}_B,\sigma_z\otimes \mathrm{I}_M \otimes \mathrm{I}_B\right),
    \label{eq:descriptorsAt0}
\end{align} where $\sigma_k$, $k=x,y,z$, are the Pauli operators and $\mathrm{I}_{i}$, $i=M,B$, is the identity operator on the qubit $Q_B$ and the unknown system $M$. Similarly:
\begin{align}
    \hat{q}^{(B)}(t_0)&\coloneqq \left(q_x^{(B)}(t_0),q_y^{(B)}(t_0),q_z^{(B)}(t_0)\right) \nonumber & \\ & =\left(\mathrm{I}_A \otimes \mathrm{I}_M \otimes \sigma_x,  \mathrm{I}_A \otimes \mathrm{I}_M \otimes \sigma_y, \mathrm{I}_A \otimes \mathrm{I}_M \otimes \sigma_z\right),
    \label{eq:descriptorsBt0}
\end{align} is the vector of descriptors of the qubit $Q_B$. The descriptors allow us to keep track of the evolution of each qubit's \textit{local algebra of observables} while acting on the whole Hilbert space. Notice that $\left[q_k^{(i)},q_l^{(j)}\right]=0$, $\forall \, i\neq j,\, k,l=x,y,z$, which implies that quantum observables of two different, non-interacting subsystems must commute.} 

{With this picture in mind, let us now prove the GWT in quantum theory. We recall that being this witness a \textit{sufficient} condition for the non-classicality of the mediator $M$, we shall assume that at the end of the protocol the two quantum probes $A$ and $B$ end up entangled.}

{Considering to the setup shown in Fig.\ref{fig:BMV}, let us assume $M$ to be a classical system. Its vector of descriptors will thus have a single component, 
\begin{equation}
    q_z^{(M)}(t_0)\coloneqq \mathrm{I}_A\otimes \sigma_z \otimes \mathrm{I}_B,
    \label{eq:descriptorsMt0}
\end{equation}
following our definition of classicality, to reflect the existence of a single observable for it. We shall identify this observable with the Pauli $Z$ operator.
At time $t$, the quantum probe $A$ interacts with the mediator $M$. The most general state of the system $A\oplus M$ is:
\begin{align}
     \rho(t)=\frac{1}{4}&\left(\mathrm{I}+\Vec{r}_A\cdot\hat{q}^{(A)}(t)+ s_zq_z^{(M)}(t) \right. \nonumber & \\ & \left. \;\; +\Vec{t}_A\cdot\hat{q}^{(A)}(t)q_z^{(M)}(t)\right),
    \label{eq:generalstate}
\end{align}
where $\Vec{r}_A$ and $\Vec{t}_A$ are real-valued vector, $s_z\in \mathbb{R}$, and $\hat{q}^{(A)}(t)$ and $q_z^{(M)}(t)$ are functions of $\hat{q}^{(A)}(t_0)$ \eqref{eq:descriptorsAt0} and $q_z^{(M)}(t_0)$ \eqref{eq:descriptorsMt0} only, respectively. This state, interpreted as a two-qubit state, is \textit{separable}, meaning that no quantum correlations can be generated between the quantum probe $A$ and a classical mediator $M$. Thus, it would be impossible to find entanglement between $A$ and $B$ at the end of the protocol as the final state would be separable too, contradicting the assumption of the witness. The reason why $\rho(t)$ in \eqref{eq:generalstate} is separable lies in the classicality of $M$: the mediator needs (at least) another observable that does not commute with $\sigma_z$. Thus, observing entanglement between $A$ and $B$ at the end of a protocol performed \textit{by local means only} leads to the conclusion that $M$ must be non-classical, according to our definition of non-classicality.}

{But why is the locality of the protocol so crucial? 
The formalism of descriptors in quantum theory allows us to understand immediately the consequences of dropping the locality assumption. 
Let us modify the GWT setup as shown in Fig.\ref{fig:BMV} and assume that the quantum probes $A$ and $B$ can interact directly in some way. This means that the descriptors of the quantum probe $A$ at time $t$ will become a function of the descriptors of quantum probe $B$ at the previous time $t_0$ in \eqref{eq:descriptorsAt0} and \eqref{eq:descriptorsBt0}, i.e., $\hat{q}^{(A)}(t)=f(\hat{q}^{(A)}(t_0),\hat{q}^{(B)}(t_0))$. If this is the case, then the state in \eqref{eq:generalstate} \textit{can} result in an entangled state for $A$ and $B$ at the end of the protocol even with a classical system $M$ as mediator. Thus, observing the entanglement between the quantum probes will \textit{not} lead to the conclusion that $M$ must be non-classical: there would be no contradiction between observing $A$ and $B$ entangled and the mediator $M$ having only a single variable. In fact, the entanglement in question could have been generated already before $A$'s interaction with $M$, irrespectively of the mediator.}

{Thus, the assumption of locality is crucial in this example to ``force" $M$ in using (at least) two non-commuting variables to entangle $A$ and $B$: one is used to entangle with $A$, the other to transmit the quantum correlations between the two quantum probes.} {However, an effective witness of non-classicality must be formulated without relying on the formalism of quantum theory, in order to consider the possibility that the unknown system $M$ may also be described by postquantum theories.} 

As we shall now explain, the locality assumption can be stated clearly as a general principle, without relying on a specific dynamics, and distinguished carefully from other notions of locality, some of which are dynamics-dependent.

\section{Locality, no-signalling, microcausality and Lorentz-Covariance}

Here we distinguish the principle of locality from other principles, such as no-signalling, and properties of specific theories, such as microcausality and Lorentz-covariance. We shall follow an order of generality: locality is the most general property, Lorentz-covariance the least general. We shall refer, for convenience, to a bipartite system, made of two subsystems: $A$ and $B$. 
\\

\textbf{Principle of Locality}.-- The \textit{principle of locality} states that given a partition of a system into subsystems $A$ and $B$, a dynamical transformation that operates only on $A$ cannot change the \textit{states} of $B$. 

Note that here by ``state" we mean the \textit{complete} specification of the state of affairs of a given system, not necessarily what is empirically accessible by measuring observables of that system only. This is also known in the literature as the ``ontic state", or ``noumenal state", to differentiate it from the ``phenomenal state", or ``epistemic state", indicating what is observable in the system, \cite{spekkens_evidence_2007,raymond-robichaud_local-realistic_2021}.
Thus, the principle of locality can also formally be stated as a strict constraint on the states of systems, as follows, \cite{deutsch_constructor_2015}:
\begin{teorema}
    The state of a system is a description of it that satisfies two properties: (i) any attribute of a system, at any given time $t$, is a fixed function of the system state and (ii) any state of a composite system $A\oplus B$ is an ordered pair of states $(a,b)$ of $A$ and $B$, with the property that if a task is performed on $A$ only, then the state $b$ of the substrate $B$ is not changed thereby.
    \label{princ:loc}
\end{teorema}

In quantum theory, the principle of locality is also called \textit{Einstein's locality}. Often times it is  confused with a different notion of locality called \textit{Bell's locality}. Bell's locality refers to the possibility of describing a set of data with a local hidden-variable theory, expressed as a stochastic, real-valued theory, thus based on \textit{c-numbers}. When Bell's inequalities are violated in a given experiment, one can therefore rule out out a c-number-based, stochastic description of reality, that is local in the sense of Einstein's locality. Quantum theory is Bell non-local; however, it is a  \textit{q-number-based}, deterministic description of reality that satisfies the principle of locality (as stated earlier). Thus, Bell's non-locality of quantum theory should not be misunderstood as a violation of the principle of locality as stated above, but rather as a violation of Bell's locality. From now on, when referring to ``locality", we will always mean the property required by the locality principle stated in Principle \ref{princ:loc}, not Bell's locality.
\\

\textbf{Principle of No-Signalling}.-- The \textit{principle of no-signalling} states that a dynamical transformation that operates only on one subsystem $A$ cannot change the {\sl observable properties} of the other subsystem $B$. 

In quantum theory, one can express this principle as follows:
\begin{equation}
    [U_A,\rho_B]=0
\end{equation}
where $U_A$ is a unitary transformation happening on the subsystem $A$ alone, while $\rho_B=\text{Tr}_A(\rho_{AB})$ is the reduced density operator of the subsystem $B$, which can be expressed as a linear combination of the generators of its observables algebra.

For example, if one considers $B$ to be a qubit, then $\rho_B=\frac{1}{2}\left(\mathbb{I}+\alpha X_B+\beta Y_B + \gamma Z_B \right)$, where $X_B$, $Y_B$ and $Z_B$ are the Pauli operators and $\mathbb{I}$ is the identity operator, generators of the qubit's observables algebra. Here, $\alpha,\beta,\gamma \in \mathbb{C}$ with $|\alpha|^2+|\beta|^2+|\gamma|^2=1$.

The principle of no-signalling and the principle of locality are related.
When considering a theory with 1:1 dynamics, including quantum theory, a general theorem shows that the two statements are equivalent, \cite{raymond-robichaud_local-realistic_2021}. This is because the 1:1 dynamics determines the \textit{completeness} of the theory, which then allows for the description of the complete state of affairs of the system (the ontic state), as required by the principle of locality, starting from its observable properties (the phenomenal state), as mandated by the no-signalling principle.
In general, while the principle of locality implies the principle of no-signalling, the converse is not true. This is because the principle of no-signalling only focuses on the locally empirically accessible features of a system. An example of a theory that satisfies no-signalling, but is non-local, is Bohmian Mechanics, \cite{goldstein_bohmian_2024}. This fact becomes important for the witness of non-classicality, which, in its most general form \cite{marletto_witnessing_2020}, relies on the principle of locality, not on the principle of no-signalling. 

Notice also that the above two principles concern notions such as ``observables", ``variables" and ``dynamical transformations", which can be expressed in many different formalisms. Hence, while these principles are both satisfied by non-relativistic quantum theory, quantum field theory, and special and general relativity, they are formulated \textit{independently of} any specific formalism. This is the power of principles as general rules that can constrain particular dynamics, rather than being derived within a particular dynamical law. 

Finally, note that the principle of no-signalling can also be phrased, within specific dynamical laws, as requiring a finite speed of propagation for signals between space-like separated systems. Here we have opted for a formulation that is dynamics-independent and does not explicitly refer to speed of propagation. 
\\

\textbf{Axiom of Microcausality}.-- In quantum field theory, the property of \textit{microcausality} is often taken as an \textit{axiom}. This property states that the allowed quantum observables (operators) of space-like separated systems must commute. In other words, 

\begin{equation}
\left[\hat{Q}_A, \hat Q_{B}\right] = 0, \label{FI}
\end{equation}
for any two quantum observables $\hat{Q}_A$ and $\hat{Q}_B$ respectively pertaining to two space-like separated systems $A$ and $B$. Note that when we talk about space-like separated systems we implicitly refer to a \textit{particular formalism}, that of special relativity. Moreover, by using the commutator, we are committing to the formalism of quantum theory. In passing, we note that equation \eqref{FI} is a property of non-relativistic quantum theory, where the quantum observables of independent subsystems are required to commute with each other. This property implies that both non-relativistic quantum theory and quantum field theory satisfy the principle of locality as stated above: since $\hat{Q}_A$ commutes with $\hat{Q}_B$ and the two theories are complete, operating a dynamical transformation on the observable $\hat{Q}_B$ of the subsystem $B$ cannot modify the observable $\hat{Q}_A$ of the subsystem $A$, i.e., its state, as required by the principle of locality.
Finally, microcausality as stated above is \textit{not} a general principle, because it is formulated within quantum theory's and special relativity's formalism. 
\\

\textbf{Axiom of Lorentz-Covariance}.-- In special relativity, allowed dynamical variables must satisfy the \textit{axiom of Lorentz-covariance}. This means that they must transform properly under Lorentz transformations so that physical variables must be either scalars, tensors, or spinors. Dynamical laws are then said \textit{Lorentz-covariant} if they are expressed in terms of Lorentz-covariant variables only. A dynamical law with this property satisfies the principle of relativity, that all laws must make the same predictions about identical experiments in any two inertial frames. Some physical properties such as scalars are also \textit{Lorentz invariant}, which means that they take the same form in all reference frames. When a law satisfies this property it must also satisfy the principles of locality and no-signalling, as the maximum speed of propagation for dynamical perturbations is the speed of light. The two principles as formulated above refer to ``instantaneous" modification of the state or the observable of the other subsystems and do not refer to any speed of propagation or other conditions in space and time. Similar concepts can be defined in general relativity, where one talks about \textit{local Lorentz-covariance} (local in spacetime). The same considerations about the relation with the principle of locality and no-signalling hold in that case. 

Once more, Lorentz-covariance is a dynamics-dependent formal requirement, which only holds within special relativity. It is implied by microcausality, but it is less general than the locality or the no-signalling principle as formulated earlier.
\\

\textbf{Universality}.--  There are also more specific notions of locality, related to the particular structure of dynamical interactions. For instance, there is the property that a given unitary on an $n$-partite system can be approximated arbitrarily well by a set of unitaries operating on one and two systems. In the field of quantum information, this is called the \textit{universality of two- and one-qubit gates}, which was proven by a sequence of results in the eighties (see \cite{divincenzo_two-bit_1995,deutsch_universality_1997, barenco_universal_1997}). 
That property can also be proven for continuous dynamics, considering that the latter can be approximated arbitrarily well by the dynamics of discrete systems. These well-known results were also recently termed ``subsystem locality", \cite{christodoulou_possibility_2019}. These properties are, once more, formalism-specific and thus more narrow than the principles discussed earlier.

\section{The role of locality in the General Witness Theorem}

According to the most general proof currently known, locality is one of the two sufficient conditions for the General Witness Theorem, \cite{marletto_witnessing_2020}. All viable theories currently trusted satisfy this principle: in particular, it applies to both special and general relativity.

Let us first state the notion of non-classicality that the witness is meant to assess, in constructor-theoretic terms.  We need to recall constructor theory's definition of a superinformation medium, which generalises the concept of a quantum system without relying on quantum theory's formalism.

Central to this definition is that of information variable. An \textbf{information variable} $X$ is a set of disjoint \textit{attributes} $\{ {\bf x} \}$ (i.e., a set of all the states where the system has a given property) for which the following tasks are possible:
\begin{align} 
    &\bigcup_{x\in X}\left\{\left({\bf x},{\bf x_0}\right)\rightarrow \left({\bf x},{\bf x}\right)\right\},  \label{eq:taskcopy} \\
    &\bigcup_{x\in X}\left\{{\bf x}\rightarrow \Pi \left({\bf x}\right)\right\}. \label{eq:taskpermutation}
\end{align}
The ``copying" task \eqref{eq:taskcopy} corresponds to copying the attributes ${\bf x} \in X$ of the first replica of the system onto the second, target, system prepared in a blank attribute $ {\bf x_0} \in X$. The ``permutation" task \eqref{eq:taskpermutation} describes the possibility of performing a logically reversible computation on the variable $X$, described as a permutation $\Pi$ on the set of labels of the attributes in $X$.
Together, the two tasks \eqref{eq:taskcopy} and \eqref{eq:taskpermutation} indicate that it is possible to perform information processing on the variable $X$. An \textbf{information medium} is a system having at least one information variable.

A \textbf{superinformation medium} is an information medium with at least two information variables, whose union is not an information variable. This means that the two information variables are complementary to one another, and cannot be copied by the same device. In \cite{deutsch_constructor_2015}, it is shown that superinformation media have all the qualitative properties of quantum systems. In this framework, it is natural and elegant to express the idea of non-classicality that is relevant to the GWT, as follows. 

By a system being {\bf non-classical} one means an information medium ${\bf M}$, with maximal information observable $Z$, that has also another variable $V$ \textit{disjoint} from $Z$ and with the \textit{same cardinality} as $Z$, with these properties:

\begin{enumerate}
\item{} There exists a superinformation medium ${\bf S_1}$ and a distinguishable variable $E=\{{\bf e_j} \}$ of the joint substrate ${\bf S_1}\oplus {\bf M}$, whose attributes ${\bf e_j}=\{(s_j, v_j)\}$ are sets of ordered pairs of states, where  $v_j$ is a state belonging to some attribute in V and $s_j$ is a state of ${\bf S_1}$;
\item{} The union of $V$ with $Z$ is {\sl not} a distinguishable variable;
\item{} The task of distinguishing the variable $E=\{{\bf e_j} \}$ is possible by measuring incompatible observables of a {\sl composite superinformation medium} including ${\bf S_1}$, but impossible by measuring observables of ${\bf S_1}$ only.
\end{enumerate}

The above conditions express the fact that a non-classical system has a classical variable (the information variable) that is also an observable; and then it has another dynamical variable that is necessary to mediate the generation of entanglement, which may or may not be directly measurable -- its attributes may or may not be all preparable or single-shot distinguishable. 

The proof of the GWT goes as follows. 
First, one considers three systems: $A$ and $B$ (two quantum probes), and $M$, the mediator, which has a ``classical" observable $Z$, and no other known degrees of freedom. In the case of gravity, the classical basis of the gravitational field would be the number operator or its energy. Let A's descriptors be denoted by ${\bf q_A(t_0)}$, B's descriptors by ${\bf q_B(t_0)}$, and the mediator's descriptors by ${\bf c(t_0)}$, \cite{deutsch_information_2000}. Let us assume that the observables (or measurable properties) of the three systems at time $t$, as well as their joint observables, are a \textit{fixed function} of the triplet $({\bf q_A(t_0)}, {\bf c(t_0)},{\bf q_B(t_0)})$. In quantum theory, this is given by the trace with the initial state's density operator, which never changes and therefore can be incorporated into the definition of the function. 

Assume that initially (at time $t=t_0$) the three systems are \textit{uncorrelated}, for instance (in quantum theory) we would say that they are in a product state where a local observable of each subsystem is sharp with some value. So the mediator will have $Z$ sharp with value say $z_0$. We assume that it is possible to run the experiment with two distinguishable initial conditions, say for instance: the case where the two masses are prepared in some state $s_{+}$, in which case the descriptors shall be labelled as ${\bf q_{A,B}^{+}(t_0)}$, and in another distinguishable state $s_{-}$, with descriptors ${\bf q_{A,B}^{-}(t_0)}$. 

Suppose that at time $t_2$ $A$ and $B$ end up in both cases being entangled. This means that at time $t_2$ they have been prepared in one of two entangled states, call them $\textbf{e}_{+}$ and $\textbf{e}_{-}$, which are distinguishable too by measuring observables of $A$ and $B$ only. Thus, the following task must be possible:
\begin{equation}
T_E \doteq \left\{\left({\bf q_A^{\pm}(t_0)}, {\bf c(t_0)}, {\bf q_B^{\pm}(t_0)} \right) \rightarrow {\bf e_{\pm}} \right\},
\label{eq:finaltask}
\end{equation} and the goal of the experiments proposed in \cite{bose_spin_2017,marletto_gravitationally_2017} is to show that $T_E$ is indeed possible upon successfully generating entanglement between $A$ and $B$.
Locality here can be used to conclude that in both configurations, the descriptors of $A$, $M$ and $B$ are an ordered triplet:
\begin{equation}
\textbf{e}_{\pm} \equiv \left({\bf q_A^{\pm}(t_2)}, {\bf c^{\pm}(t_2)},{\bf q^{\pm}_B(t_2)}\right).
\end{equation}

Now, we can use the assumption that $A$ and $B$ are not interacting directly, but the interaction is mediated by $M$. To a first approximation, this means that the interaction happens by letting first $A$ interact with the mediator $M$, at time $t_1$, and then $M$ interact with the qubit $B$, at time $t_2$. 
In terms of tasks, $T_E$ in \eqref{eq:finaltask} must thus be made by:
\begin{equation}
    T_E^{(1)}\doteq \left\{\left({\bf q_A^{\pm}(t_0)}, {\bf c(t_0)}, {\bf q_B^{\pm}(t_0)} \right) \rightarrow {\bf r_{\pm}} \right\},
\end{equation} performed on $A\oplus M$ only, and:
\begin{equation}
    T_E^{(2)}\doteq \left\{{\bf r_{\pm}} \rightarrow {\bf e_{\pm}} \right\},
\end{equation} performed on $M\oplus B$ only.
Once more using locality, the state of the three systems at time $t_1$ must be described by one of two triplets, according to whether the initial condition $s_+$ or the initial condition $s_-$ was used:
\begin{equation}
\textbf{r}_{\pm} \equiv \left({\bf q_A^{\pm}(t_1)}, {\bf c^{\pm}(t_1)},{\bf q^{\pm}_B(t_0)}\right).
\end{equation}
Locality is here used because we have considered that the descriptor of system $B$ must not have changed since $t_0$, given that the interaction at $t_1$ only involves $A$ and $M$.

The proof proceeds by showing that the descriptors ${\bf c^{\pm}(t_1)}$ are: 1) disjoint (set-wise) from one another, and from the classical states of $M$; 2) that they are not distinguishable from the classical states of $M$; 3) that they are not distinguishable from one another, yet the joint state of $A$ and $M$ is (using measurements that involve both $A$ and $M$). 
These three properties make $M$ non-classical, in that it has a dynamical variable $V$, made of the two descriptors $\{{\bf c^{\pm}(t_1)}\}$ which is disjoint from the classical observable, and yet is not distinguishable from the latter, just like Heisenberg's uncertainty principle requires. We refer the interested reader to \cite{marletto_witnessing_2020} for detailed proof of the above three points.

As we mentioned earlier, it is important to notice that the non-classical variable $V$ of the mediator, unlike $Z$, may not be an observable, in the sense that ${\bf c^{+}(t_1)}$ may not be distinguishable from ${\bf c^{-}(t_1)}$ in a single shot manner, and that those two states may not be preparable. 

It follows from this argument that violating the assumption of locality \textit{invalidates} the witness, giving the appearance that even a classical mediator $M$ can create entanglement between the two quantum probes $A$ and $B$. This is the case of models using the non-local Newtonian gravity, \cite{marchese_newtons_2024}, allowing for interactions between $A$ and $B$, \cite{husain_dynamics_2022}, or assuming a simultaneous interaction of the mediator $M$ with both the quantum probes, \cite{martin-martinez_what_2023}. Moreover, some quantum-classical hybrid models, \cite{hall_two_2018}, conceal hidden non-locality in the configuration space dynamics, as noted in \cite{marconato_vindication_2021}.
Violating locality means removing the assumption involving the second, non-compatible variable $V$ of $M$ in creating entanglement, as shown in the proof above. This is because the task $T_E$ would be performed in a single step, instead of at least two. Thus, with a non-local model, $M$ would need a single variable to perform $T_E$, ultimately a classical system. However, as discussed above, both quantum mechanics and general relativity obey the principle of locality in the general form introduced in this work, thus securing its presence among the minimal assumptions of a theorem aimed at proving the non-classical nature of an unknown system. 

\section{Conclusions}

The principle of locality as discussed in this work is a plausible candidate for the most general notion of locality in physics, considering its relation to the other relevant notions that we have reviewed in this paper. It is also the minimal notion of locality on which the witness of non-classicality for the BMV effect \cite{marletto_witnessing_2020} must rely. 

One interesting open question is in the direction of considering a temporal equivalent of the GWT, with a possible notion of locality in time, not just in space.

Following the pioneering work by Leggett and Garg, \cite{leggett_quantum_1985}, which demonstrated that quantum systems exhibit a form of temporal correlation that any macro-realistic theories cannot explain, the study of temporal correlations and their relationship to spatial correlations has become increasingly popular in the scientific community, \cite{emary_leggettgarg_2013}. Central to this topic is the idea of \textit{locality in time}, which was explored in \cite{brukner_quantum_2004} and formulated as follows: the results of measurement performed at time $t_2$ are independent of any measurement performed at some earlier or later time $t_1$. 
Despite this formulation, a consensus on the precise meaning of locality in time remains elusive, and it is still debated whether quantum mechanics should be interpreted as adhering to temporal locality \cite{adlam_spooky_2018, cohen_quantum_2023}. It is therefore an interesting open question how to relate it to the principle of locality ``in space", as discussed earlier in this paper.

A recently proposed witness of non-classicality, \cite{di_pietra_temporal_2023} provides a new framework for probing the quantum nature of an unknown system $M$ by studying the time evolution of a single quantum probe $Q$ -- thus providing a ``temporal equivalent" of the entanglement-based witness just discussed. Specifically, if a system $M$ induces a quantum coherent evolution in a quantum probe $Q$ while conserving a global quantity of the system $Q\oplus M$, then $M$ must be quantum. This ``temporal" witness mirrors the spatial witness of non-classicality proposed in \cite{bose_spin_2017, marletto_gravitationally_2017}, where the ability of $M$ to generate non-classical spatial correlations between two probes $A$ and $B$ is examined. In contrast, the temporal witness focuses on detecting non-classical temporal correlations in the evolution of the quantum probe $Q$.

Given the crucial role played by the assumption of locality in space in the GWT, it is reasonable to expect that locality in time may play an analogous role in the temporal witness. The connections between these two witnesses suggest that the assumptions underpinning each may be closely related: while locality in space is central to the GWT, the conservation law of a global quantity in the system $Q\oplus M$ could be linked to locality in time in the temporal witness. This raises the intriguing possibility that the conservation law required for the quantum coherent evolution of $Q$ is not only essential for detecting non-classicality but may also offer insights into the nature of temporal locality.

In this direction, following those attempts in quantum theory to treat space and time on equal footing, \cite{zhang_different_2020}, one could interpret the two subsystems $A$ and $B$ in the spatial witness as a \textit{single} system at \textit{two different times} $t_A$ and $t_B$ in the temporal one, and follow the general proof described in the previous section. Can the principle of locality as stated in this paper be applied also in this scenario? Where does the conservation law play a role in the general proof for the temporal case? 

We leave the answers to these questions to future work.
\\

{\bf Acknowledgements} \;\; G.D.P. thanks the Clarendon Fund and the Oxford-Thatcher Graduate Scholarship for supporting this research. This research was made possible through the generous support of the Gordon and Betty Moore Foundation, the Eutopia Foundation and of the ID 62312 grant from the John Templeton Foundation, as part of the \href{https://www.templeton.org/grant/the-quantuminformation-structure-ofspacetime-qiss-second-phase}{‘The Quantum Information Structure of Spacetime’ Project (QISS)}. The opinions expressed in this project/publication are those of the author(s) and do not necessarily reflect the views of the John Templeton Foundation. 

\bibliographystyle{apsrev4-2}
\bibliography{locality}

\end{document}